\documentclass[journal]{rmaa}

\usepackage{paralist}

\usepackage{psfrag,color}

\usepackage[latin1]{inputenc}
\usepackage{epsfig}
\usepackage{amsmath}
\usepackage{graphicx}
\usepackage{subfig}
\usepackage{multirow}
\usepackage{array}
\usepackage{rotating}

\title{ROTSE1 J164341.65+251748.1: \\ a new W UMa-type eclipsing binary} 

\author{
R. Michel\altaffilmark{1}, 
J. Echevarr\'ia\altaffilmark{2}, 
T.Q. Cang\altaffilmark{3}, 
L. Fox-Machado\altaffilmark{1} 
and D. Gonz\'alez-Buitrago\altaffilmark{1}}

\altaffiltext{1}{
Instituto de Astronom\'ia, 
Universidad Nacional Aut\'onoma de M\'exico, 
Apartado Postal 877, 
Ensenada, Baja California, C.P. 22830 M\'exico. e-mail: rmm@astro.unam.mx}
\altaffiltext{2}{
Instituto de Astronom\'ia, 
Universidad Nacional Aut\'onoma de M\'exico, 
Apartado Postal 70-264, Ciudad Universitaria, 
M\'exico D.F., C.P. 04510, M\'exico.}
\altaffiltext{3}{
Department of Astronomy, 
Beijing Normal University, 
100875, Beijing, China}

\shortauthor{R. Michel et al.}
\shorttitle{ROTSE1 J1643+25: a new W UMa-type eclipsing binary}

\fulladdresses{
\item R. Michel: Universidad Nacional Aut\'onoma de M\'exico. 
Observatorio Astron\'omico Nacional. Apartado Postal 877, C.P. 22800, Ensenada, B.C., M\'exico.
  (rmm@astro.unam.mx.)
}
\listofauthors{R. Michel, J. Echevarr\'ia, T.Q. Cang, L. Fox-Machado \& D. Gonz\'alez-Buitrago}
\indexauthor{Michel, R.}
\indexauthor{Echevarr\'ia J.}
\indexauthor{Cang T.Q.}
\indexauthor{Fox-Machado L.}
\indexauthor{Gonz\'alez-Buitrago D.}

\abstract{

ROTSE1 J164341.65+251748.1 was photometrically observed in the V band during three epochs with the 0.84-m telescope of the San Pedro M\'artir Observatory in Mexico. Based on additional BVR photometry, we find that the primary star has a spectral type around G0V. The light curve of the system is typical of a W~UMa type binary stars and has an orbital period of $\sim$ 0.323 days. In an effort to gain a better understanding of the binary system and determine its physical properties, we analyzed the light curve with the Wilson and Devinney method. We found that ROTSE1 J164341.65+251748.1 has a mass ratio of $\sim$ 0.34 and that the less massive component is over 230 K hotter than the primary star. The inclination of the system is $\sim$ 84.6 degrees, and the {\bf degree} of over-contact is 11\%. The analysis shows the presence of variable bright spots on the primary star.}

\resumen{

ROTSE1 J164341.65+251748.1 fue observada fotom\'etricamente en la banda V durante tres \'epocas con el telescopio de 0.84-m del Observatorio de San Pedro M\'artir en M\'exico. Con base en fotometr\'ia BVR adicional, encontramos que la estrella primaria tiene un tipo espectral cercano a G0V. La curva del luz del sistema es caracter\'istica de los sistemas binarios del tipo W~UMa y tiene un periodo orbital de $\sim$ 0.323 d\'ias. Con el prop\'osito de entender mejor a este sistema binario, hemos analizado su curva de luz con ayuda del m\'etodo de Wilson y Devinney. Hemos encontrado que ROTSE1 J164341.65+251748.1 tiene una raz\'on de masas de $\sim$ 0.34 y que la componente menos masiva es hasta 230 K m\'as caliente que la estrella primaria. La inclinaci\'on del sistema es $\sim$ 84.6 grados y el sobre-contacto es de 11\%. El an\'alisis tambi\'en muestra la presencia de manchas brillantes variables sobre la estrella primaria.}

\addkeyword{techniques: photometric}
\addkeyword{binaries: eclipsing}
\addkeyword{stars: individual: ROTSE1 J164341.65+251748.1}

\begin{document}
% Typeset article header
\maketitle

\section{Introduction}
\label{sec:intro}

The ROTSE-I survey for variable stars \citep{ake00} has identified nearly 2000 new variables. Among them {\bf is} ROTSE1 J164341.65+251748.1 (hereinafter J1643+2517), reported by these authors as a variable source with a V magnitude of around 14.2, varying with an amplitude of 0.69 mag and a periodicity of 0.32343200 days. No further information is given about the nature of its variability. While conducting a spectroscopic and photometric study in the V band on AH Herculis, J1643+2517 was detected as a serendipitous source in the field.  Our extensive photometric coverage reveals that J1643+2517 has the typical light curve of W~UMa-type systems. 

Since the pioneer work by \citet{luc68} on contact stars, it is now well established that W~UMa-type systems are over-contact binary stars, which consist of two late type main sequence stars sharing a common connecting envelope \citep{ruc10}. There is a considerable number of publications about this kind of objects, both observational and theoretical, since the early work by \citet{ege61}, who recognized that binaries with orbital periods of less than a day and with continuous light variations, could be considered stars in full contact. Among these are the reviews published by \citet {ruc93} and \citet{egg96}. In this paper, based on the obtained V light curves of J1643+2517, we determine the most important parameters of the system.

\section{Observations and Data Reduction}
\label{sec:observations}

CCD observations were made during three observing runs, with the 0.84-m f/15 Ritchey-Chr\'etien telescope at Observatorio Astron\'omico Nacional at Sierra San Pedro M\'artir (OAN-SPM), the Mexman filter-wheel, and the ESOPO CCD detector (e2v CCD42-90), which has a 2048$\times$4608 13.5 $\mu$m square pixel array, a gain of 1.7 e$^-$/ADU, and a readout noise of 3.8 e$^-$. A 2$\times$2 binning was used during our observations. The combination of the telescope and detector characteristics ensured an unvignetted field of view of $7.4^{\prime}\times9.3^{\prime}$. Since the observations were obtained with AH Her as the primary target, we obtained V photometry mainly. During our last run we made BVR photometry, basically to obtain information leading to derive the spectral type of the primary. The log of the observations is shown in Table \ref{tab:phot-log}. Sky flats and bias exposures were also taken each night. 

\begin{table}
\caption{Log of photometric observations of J1643+2517.}
\label{tab:phot-log}
\begin{center}
{\tiny
\begin{tabular}{lcccc}
\hline\hline
  Date    & Filter & Exp. Time & images & Obs. Time  \\
 yyyymmdd &         & (secs)   &        & (hours)    \\
\hline
20130530 & V &  30 & 439 & 4.85 \\
20130531 & V &  30 & 709 & 8.27 \\
20130601 & V &  30 & 263 & 2.93 \\
20130602 & V &  30 & 683 & 7.87 \\
20130603 & V &  30 & 376 & 4.57 \\
20130604 & V &  30 & 281 & 3.65 \\
20130605 & V &  30 & 340 & 4.57 \\
20130606 & V &  30 & 267 & 2.90 \\
20130607 & V &  30 & 223 & 2.42 \\
20130613 & V &  30 & 603 & 7.19 \\
20130614 & V &  30 & 479 & 7.03 \\
20130617 & V &  30 & 574 & 6.64 \\
20130618 & V &  30 & 373 & 4.52 \\
\hline
20140214 & V & 180 & 43 & 3.25 \\
20140215 & V & 180 & 28 & 2.30 \\
20140216 & V &  60 & 42 & 1.526 \\
20140220 & V & 120 & 17 & 0.64 \\
20140308 & V &  20 & 532 & 4.66 \\
20140312 & V &  20 & 475 & 4.86 \\
20140519 & V &  60 & 208 & 4.26 \\
20140520 & V &  20 & 210 & 1.80 \\
\hline
20150630 & V &  30 & 264 & 3.33 \\
20150702 & B,V,R & 30,20,15 & 106,106,106 & 4.47 \\
20150703 & B,V,R & 30,20,15 & 47,47,47    & 3.13 \\
20150729 & B,V,R & 30,20,15 & 151,151,151 & 5.79 \\
20150730 & B,V,R & 30,20,15 & 51,51,51    & 2.01 \\
\hline
\end{tabular}
}
\end{center}
\end{table}

All CCD images were processed using the IRAF\footnote{IRAF is distributed by the National Optical Observatories, operated by the Association of Universities for Research in Astronomy, Inc., under cooperative agreement with the National Science Foundation.} package. Images were bias subtracted and flat field corrected and then cosmic rays were removed with the help of the L.A. Cosmic\footnote{\url{http://www.astro.yale.edu/dokkum/lacosmic}} script. Then, instrumental magnitudes of the stars were computed using the standard aperture photometry method.

\begin{figure}
\begin{center}
\includegraphics[scale=0.40]{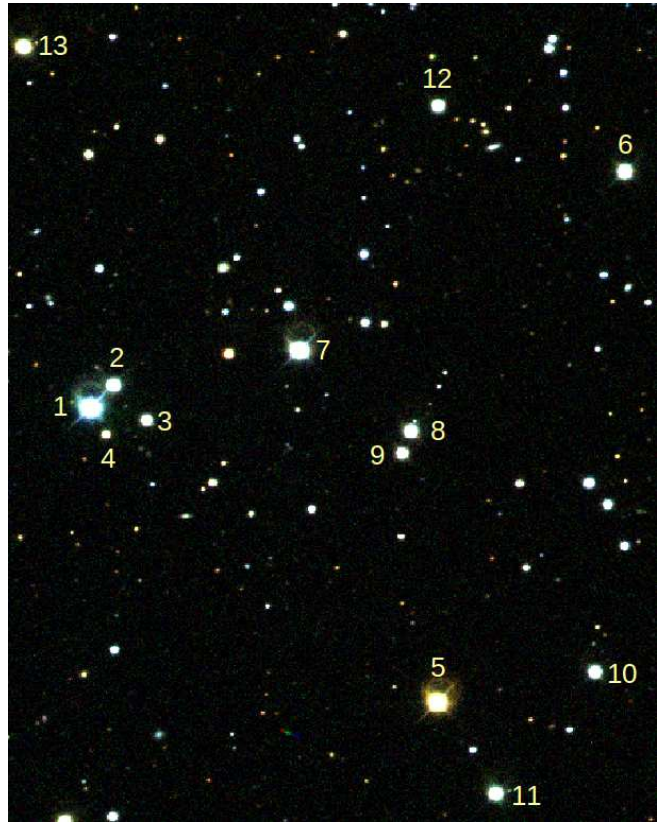}
\end{center}
\caption{Observed field. The numbered stars are calibrated photometrically and are shown in Table \ref{tab:UBVRI_mags}. J1643+2517 is star number 6 while AH Her is star number 1. The comparison star used is number 12.}
\label{fig:field}
\end{figure}

During an observing run of UBVRI photometry of Galactic stellar clusters, this field (Fig \ref{fig:field}) was included in the list of targets so that its stars could be calibrated in the Johnson-Cousins system. The results are presented in Table \ref{tab:UBVRI_mags}. Although dimmer than other stars in the field, star number 12 was used as the comparison star to obtain the differential photometry, since it has a similar colour to J1643+2517.

\begin{sidewaystable}
\caption{UBVRI photometry of the field stars. The values on the last two columns are those found in the literature.}
\label{tab:UBVRI_mags}
\begin{center}
{\scriptsize
\begin{tabular}{cccccccccccccccc}
\hline\hline
ID  & RA & DEC & U & B & V & R & I & eU & eB & eV & eR & eI & Name & B0 & V0 \\
\hline
 1 & 251.04172 & +25.25060 & 11.517 & 12.354 & 12.297 & 12.143 & 11.954 & 0.001 & 0.003 & 0.002 & 0.004 & 0.007 & V*AHHer                &        &        \\
 2 & 251.03647 & +25.25510 & 14.788 & 14.747 & 14.139 & 13.769 & 13.408 & 0.002 & 0.007 & 0.001 & 0.002 & 0.005 & [HH95]AHHer-27         & 14.800 & 14.187 \\
 3 & 251.02930 & +25.24806 & 16.202 & 15.815 & 15.058 & 14.621 & 14.238 & 0.004 & 0.005 & 0.002 & 0.002 & 0.009 & [HH95]AHHer-26         & 15.892 & 15.117 \\
 4 & 251.03827 & +25.24531 & 18.520 & 17.322 & 16.164 & 15.442 & 14.843 & 0.012 & 0.012 & 0.002 & 0.007 & 0.008 & [HH95]AHHer-25         & 17.429 & 16.205 \\
 5 & 250.96619 & +25.19135 & 15.274 & 13.851 & 12.532 & 11.831 & 11.194 & 0.003 & 0.004 & 0.002 & 0.003 & 0.004 & 2MASSJ16435186+251128  &        &        \\
 6 & 250.92379 & +25.29666 & 14.573 & 14.432 & 13.784 & 13.380 & 13.024 & 0.016 & 0.004 & 0.001 & 0.003 & 0.010 & J1643+2517             &        &        \\
 7 & 250.99569 & +25.26163 & 13.117 & 13.184 & 12.670 & 12.369 & 12.059 & 0.002 & 0.003 & 0.002 & 0.004 & 0.004 & 2MASSJ16435895+2515418 &        &        \\
 8 & 250.97129 & +25.24536 & 14.762 & 14.749 & 14.079 & 13.654 & 13.224 & 0.002 & 0.007 & 0.002 & 0.002 & 0.007 & [HH95]AHHer-21         & 14.721 & 14.035 \\
 9 & 250.97337 & +25.24101 & 15.990 & 15.580 & 14.720 & 14.234 & 13.751 & 0.001 & 0.007 & 0.002 & 0.003 & 0.006 & [HH95]AHHer-22         & 15.565 & 14.688 \\
10 & 250.93149 & +25.19695 & 15.181 & 15.029 & 14.417 & 14.040 & 13.696 & 0.001 & 0.006 & 0.002 & 0.004 & 0.005 & 2MASSJ16434356+2511489 &        &        \\
11 & 250.95348 & +25.17302 & 14.700 & 14.627 & 14.022 & 13.605 & 13.243 & 0.005 & 0.011 & 0.004 & 0.003 & 0.009 & [HH95]AHHer-18         & 14.592 & 13.938 \\
12 & 250.96468 & +25.31011 & 15.299 & 15.243 & 14.602 & 14.201 & 13.836 & 0.006 & 0.009 & 0.003 & 0.005 & 0.011 & [HH95]AHHer-33         & 15.226 & 14.560 \\
13 & 251.05568 & +25.32264 & 15.771 & 14.723 & 13.665 & 13.079 & 12.618 & 0.006 & 0.011 & 0.003 & 0.010 & 0.001 & [HH95]AHHer-2          & 14.714 & 13.645 \\
\hline
\end{tabular}
}
\end{center}
\end{sidewaystable}

\section{Photometric Analysis}
\label{sec:data}

\begin{figure*}
\begin{center}
\includegraphics[scale=0.45]{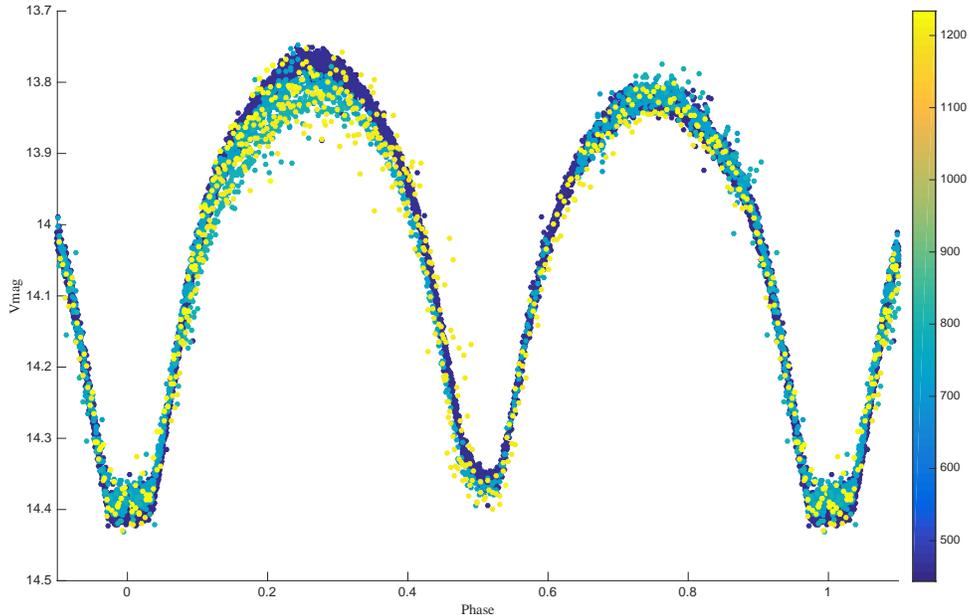}
\end{center}
\caption{Phase diagram of the V photometry of J1643+2517 (see text for the colour code). Notice that there is a clear distinction between the primary and secondary eclipses and that the primary's maxima (phase 0.25) show much larger variations that the secondary maxima (phase 0.75).}
\label{fig:phot}
\end{figure*}

The orbital period of the light curve was derived from a frequency search algorithm, using the Peranso program \citep{van14}. The derived period is $P= 0.323456 \pm 0.000024$ days. The following ephemeris has been obtained:

\begin{equation}
HJD_{\rm 0} = 2456443.5675(9)  + 0.3234561 (4) \, E,
\end{equation}

where phase zero is defined by the lower of the two minima, corresponding in this case, to the inferior conjunction of the primary star. This definition follows the convention {\bf that the deeper minimum of the light curve is defined as the primary minimum} (e.g. \citet{ bin84}). Therefore, the primary maximum occurs at phase 0.25. The parentheses indicate the error in the previous digit. Using this ephemeris, we obtain the phased light curve of J1643+2517 shown in Figure \ref{fig:phot}. The Figure shows our three observing runs. In order to identify the observing time sequence, we have included a colour bar in which the ticks are in HJD days and show the {\bf date starting from 2456443}.

\section{Spectral type of the primary star}
\label{sec:spec-types}

From the photometry obtained at phase zero, we find colours  $B-V=0.64$ and $V-R=0.43$. The $B-V$ index indicates a G2V star, while the $V-R$ index suggests a later spectral type of G9V. Thus the system appears to have a small extinction. We adopt a G0V spectral type to reconcile these indices and an initial temperature of 5900 K. 
This spectral type and  temperature are important, as we said, in the sense that they provide initial values for modeling the binary. 

\section{Modeling the Binary}
\label{sec:model}

A preliminary solution of the phased light curve was obtained by \citet{lfm15a}  with the {\scshape phoebe} package \citep{paz05}, following  the strategy explained in \citet{lfm15b}. {\scshape phoebe} is an interactive software package for modeling eclipsing binary stars based on the Wilson-Devinney code (WDc) (\citet{wad71, wil90, wil94}). It permits the creation of a synthetic light curve that fits the observational data by adjusting interactively the orbital and stellar parameters through a user friendly graphical  interface.  The preliminary results from \citet{lfm15a} were used in the present paper as start solution for the latest version of the WDc code   \citep{wav14} available in 2015 in the electronic link ftp://ftp.astro.ufl.edu/pub/wilson/lcdc2015/.
 
A final solution was then achieved using this new code. Some insights about the morphology of the binary system can be obtained from the shape of the light curve. Since there is a total eclipse of the secondary star, by the primary star, a flat bottom at the primary minimum is seen. indicating a high inclination angle of the system. The primary and secondary eclipses occur at 0.5 phase intervals, with no clear beginning and end of the eclipses suggesting a circular orbit. These properties are consistent with a short period over-contact binary of the W~UMa type. The lower minima appears at the primary minimum when the secondary star is covered by the primary star at around phase zero (see Section~\ref{sec:data}). In this case, we obtain a lower effective temperature of the more massive star. The different magnitudes of two maxima in the light curve is known as the {\it O'Connell effect}, after first being recognized by \citet{oco51}. 

During all the observations of J1643+2517, the primary maxima show strong variations, with the profile of the light curve changing in the time scale larger than a day. The secondary maxima are much more stable, particularly during the third run. To model all of these variations, we found that the best solution was one with two hot spots on the primary star. Since we cannot make a uniform model for the light curves, we separated the data into the three observed epochs in order to do their independent analyses. 

 \begin{figure*}
\begin{center}
\includegraphics[scale=0.90]{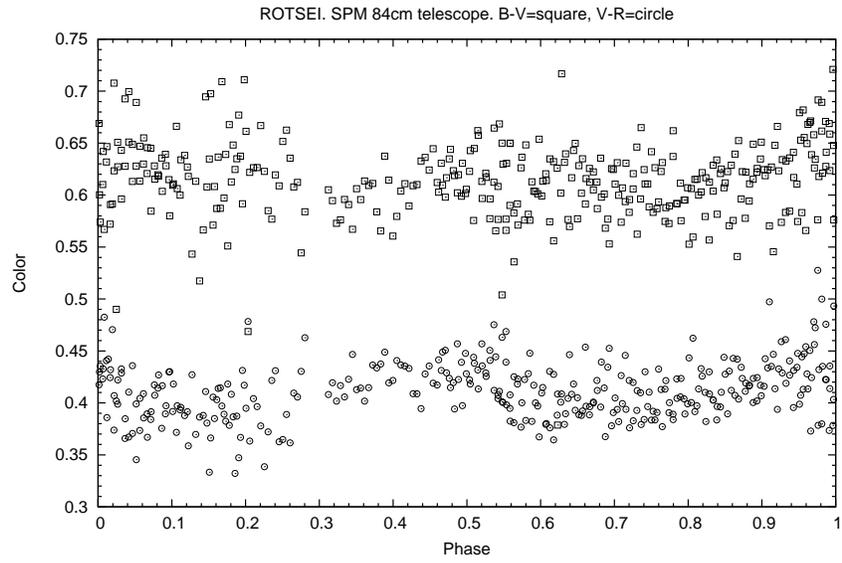}
\end{center}
\caption{The color index diagram in 2015 data. The B-V and V-R colour indexes at around 0 phase are smaller than those at 0.5 phase, which indicates a lower temperature for the primary star.  }
\label{fig:BVR}
\end{figure*}

Some of the parameters of the contact binary model could be fixed or well estimated. The thermal albedo A1 = A2 =0.5\,(\citet{luc67}) and the gravitational constant g1 = g2 =0.32 (\citet{ruc69}) are assumed fixed by the convective envelope of both stars, and the limb darkening is automatically interpolated by the WDc from the Van Hamme's table (\citet{VH1993}). Based on the BVR observations (see Section~\ref{sec:spec-types}), the spectral class can be estimated as close to G0V, giving an initial temperature of the primary of $\sim$ 5700 K. 

\begin{figure*}
\begin{center}
\includegraphics[scale=0.50]{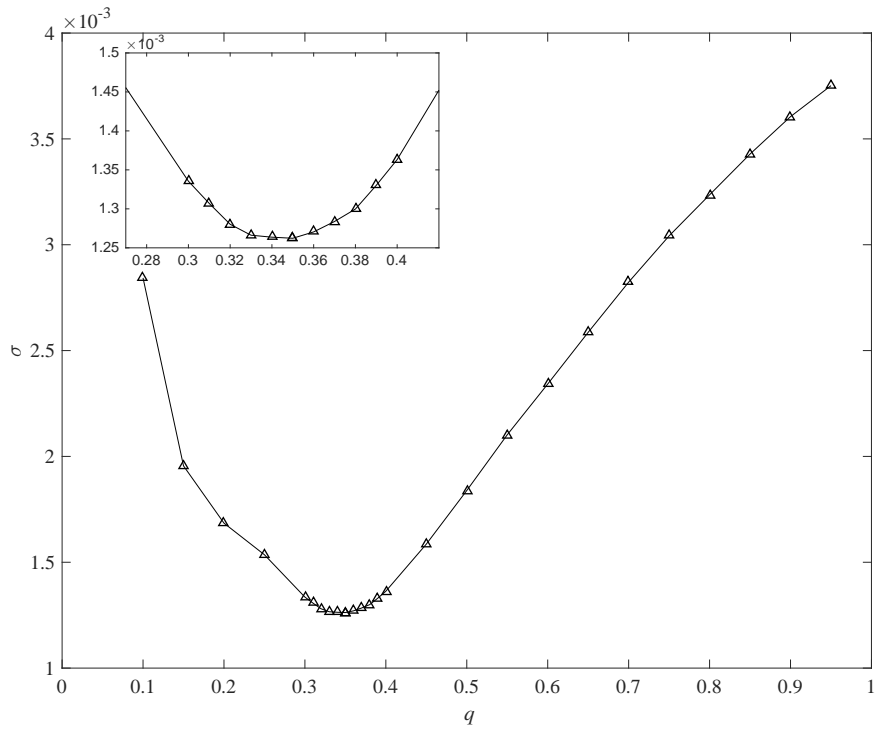}
\end{center}
\caption{The q-search strategy for the mass ratio determination. The mass ratio at minimum $\sigma$ is at around 0.35.}
\label{fig:qsearch}
\end{figure*}

\begin{figure*}
\begin{center}
\includegraphics[scale=0.4]{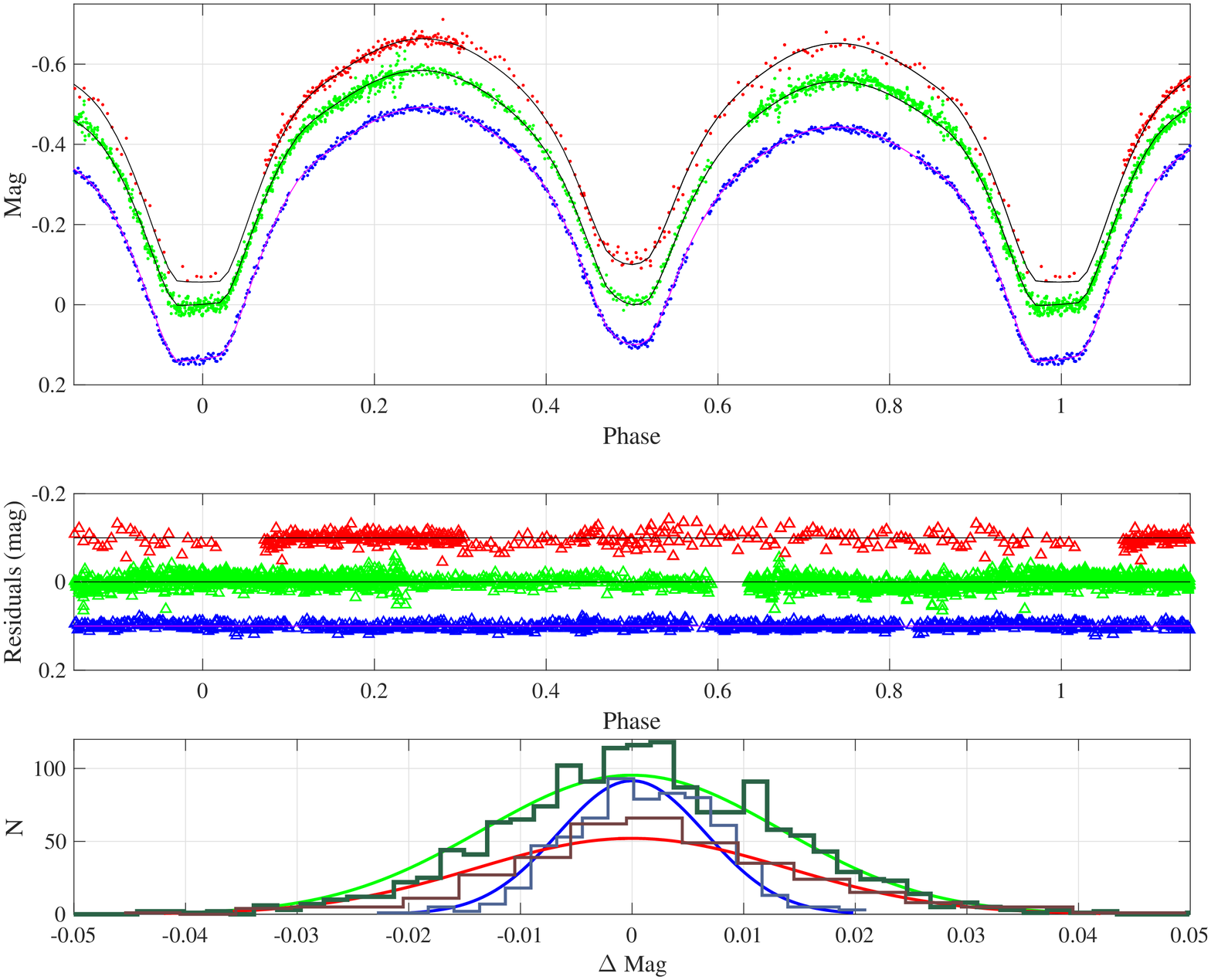}
\end{center}
\caption{$Top$. Simulated (solid lines) light curves fitted to the observations. Blue points correspond to the 2013 observations, green points to the 2014 observations and red points to 2015 observations. 
$Middle$. The residuals of the simulations. 
$Bottom$. The statistical distribution of residuals and their Gaussian distribution fits.}
\label{fig:three_model}
\end{figure*}

As the standard to build the model for this system, we used the 2013 data, which has the best photometric quality. A q-search strategy is applied to find out the mass ratio of the system (Fig. \ref{fig:qsearch}). The best fitted parameters of the simulation are listed in Table \ref{tab:Solparam}, and comparisons between simulations and observations are displayed in Fig \ref{fig:three_model}. 

\begin{figure*}
\centering
\begin{tabular}{c}
\subfloat[]{\label{}\includegraphics[width=1.0\textwidth]{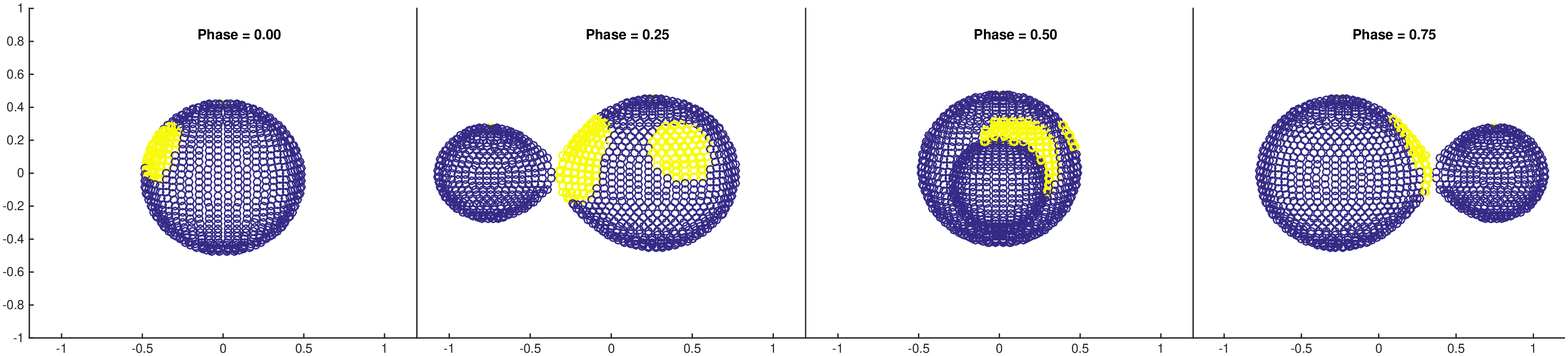}} \\ 
\subfloat[]{\label{}\includegraphics[width=1.0\textwidth]{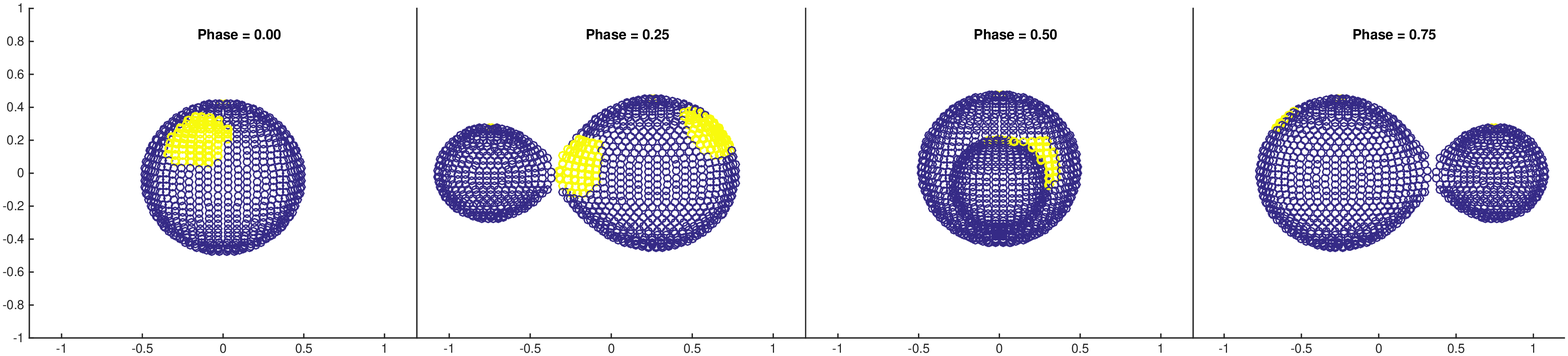}} \\ 
\subfloat[]{\label{}\includegraphics[width=1.0\textwidth]{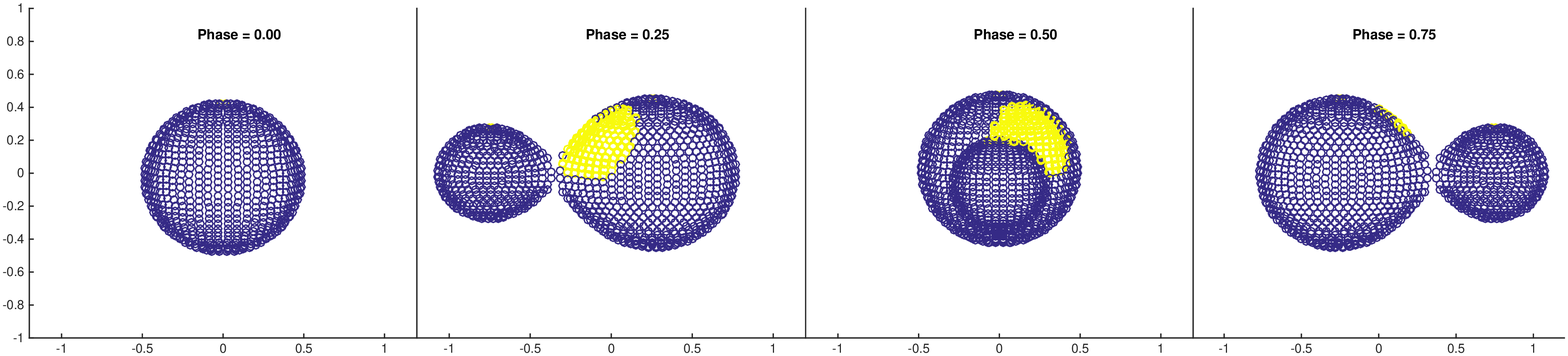}} \\ 
\end{tabular}
\caption{The shape of system for the three epoch. We present four subfigures of the shape for each epoch. The figures 
from $left-top$ to $right-bottom$ are of first, second, third year, respectively. The blue part is the shape 
of the star, and the yellow part represent the hot spots.}
\label{fig:shape_fig}
\end{figure*}

We also provide the residuals and the statistics histogram in the same figure.  With the Kolmogorov-Smirnov 
test, the residuals could be considered as a Gaussian distribution with $\sigma \sim0.0064$ at 95\% confidence. 
We also present in Fig \ref{fig:shape_fig} the shape of the system provided by the new feature of the WDc.

\section{Discussion}
\label{sec:discuss}

Our analysis of this eclipsing binary reveals properties similar in many respects to those of the W UMa 
systems [\citet{Zhu2010}; \citet{dim15}], which are characterized by having short orbital periods 
and an over-contact configuration that is composed of F-K stars sharing a common 
envelope that thermalizes the stars. This system appears to have a mass ratio of $q \approx 0.323$, 
an inclination of $i \approx 84.3^{\circ}$, and a secondary star temperature of $T_{2} \approx 5873$K. 
It is also interesting to note that, based on our results, the degree of over-contact in the system, 
$$f= \frac{\Omega_{in}  - \Omega}{\Omega_{in} - \Omega_{out}},$$
where the potentials $\Omega_{in}$ and $\Omega_{out}$ define the inner and outer critical surfaces in Roche 
geometry and $\Omega$ is the potential corresponding to the surface of the over-contact binary, is calculated 
to be $f \sim 11.5$\%. 

\begin{figure*}
\begin{center}
\includegraphics[scale=.50]{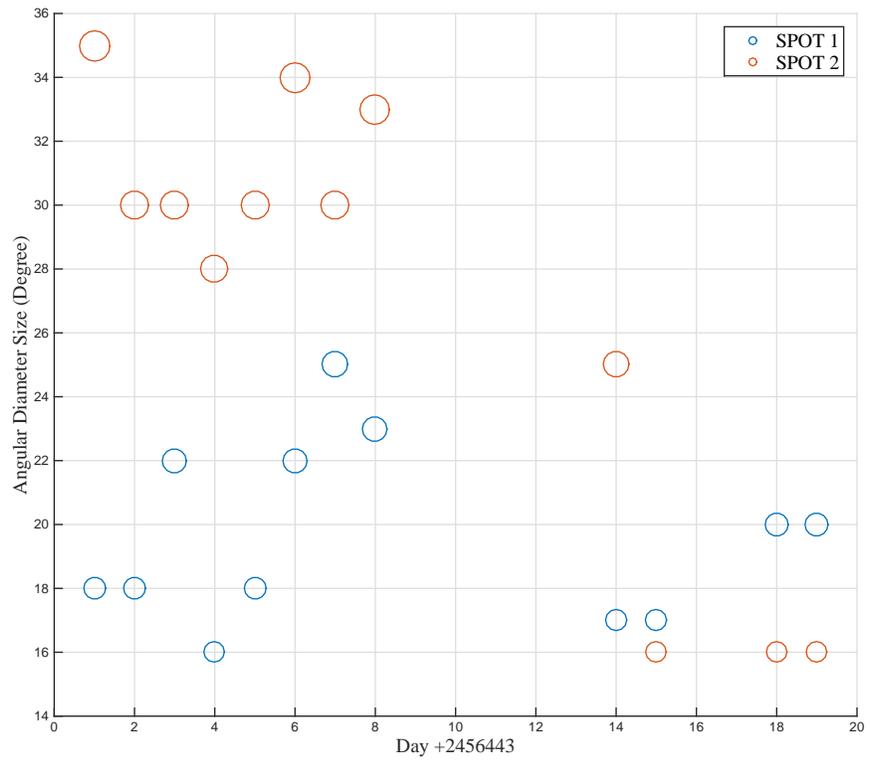}
\end{center}
\caption{The variation of spots in the first epoch. Spot 2(red) presents a significant decline in 
decades of days.}
\label{fig:spot_variation}
\end{figure*}

The fact that the primary maximum is variable and sometimes brighter than the secondary maximum can be explained by
hot spot activity. The variation of the light curve profile can be well simulated by adjusting the parameters of the spots. For a 
simplified strategy, we assumed at the beginning that the location and temperature of the spots were fixed.  
Fig \ref{fig:spot_variation} presents the size variation of two hot spots in 
19 days in the first part.  Fig \ref{fig:three_model} and Table \ref{tab:spotpara} present the spots 
parameter results of the second and third epochs, and Fig \ref{fig:shape_fig} presents the star shapes. Two similar spots can be identified  in the first and second epochs, but only one spot can be found in the third part of data. 
With all the parameters found, the temperature of the massive component is about 230 K cooler than the smaller one.

\begin{table}
\caption{Spots parameters for the data of each part. The spots are all on the primary star. There's only one 
significant spot on the system for the third part of data. }
\label{tab:spotpara}
\begin{center}
\begin{tabular}{lcc}
\hline\hline
  Parameter       &        Spot 1      &         Spot 2       \\
\hline
$Epoch 1$ \\
Colatitude (deg)  &  68.0  $\pm$ 1.3   &  75.0    $\pm$ 1.4   \\
Longitude (deg)   & 112.00 $\pm$ 0.02  &  16.00   $\pm$ 0.02  \\
Radius (deg)      &  22.0  $\pm$ 4.7   &  30.0    $\pm$ 1.4   \\
Temp. factor      &   1.07 $\pm$ 0.012 &   1.0500 $\pm$ 0.002 \\
$Epoch 2$ \\
Colatitude (deg)  &  55.0  $\pm$ 0.52  &  85.0    $\pm$ 1.46  \\
Longitude (deg)   & 155.00 $\pm$ 0.016 &  22.01   $\pm$ 0.02  \\
Radius (deg)      &  21.0  $\pm$ 2.7   &  19.0    $\pm$ 1.1   \\
Temp. factor      &   1.06 $\pm$ 0.022 &   1.0500 $\pm$ 0.015 \\
$Epoch 3$ \\
Colatitude (deg)  &                    &  60.0    $\pm$ 9.4   \\
Longitude (deg)   &                    &  34.90   $\pm$ 0.16  \\
Radius (deg)      &                    &  28.1    $\pm$ 9.48  \\
Temp. factor      &                    &   1.026  $\pm$ 0.002 \\
\hline
\end{tabular}
\end{center}
\end{table}

\begin{table*}
\caption{Light-curve parameters of J1643+2517 at 2013.06.01, the parameters with superscript ${\lowercase {a}}$ 
are assumed}
\label{tab:Solparam}
\begin{center}
\begin{tabular}{lcc}
\hline\hline
      Parameter  
      &   Primary  &  Secondary    \\
\hline
HJD$_0$                           & 2456443.5675$\pm$0.0009 \\
$P_{orb}$  (day)                   & 0.3234561$\pm$0.0000004 \\
$incl$ 	                          & 84.35 $\pm$0.31 \\
q=m2/m1                           & 0.3425 $\pm$0.0014 \\
$\Omega$=$\Omega_{1}$=$\Omega_{2}$ & 2.520 $\pm$0.005\\
T ($K$)                           & $5648 ^{{a}}$   &   5873$\pm$ 7        \\
$g1=g2$                           & $0.32^{a}$ \\
$A1=A2$                           & $0.5^{a}$ \\
$x1(bolo)=x2(bolo)$               & $0.646^{a}$ \\
$y1(bolo)=y2(bolo)$               & $0.212^{a}$ \\
$x1_{v}=x2_{v}$                    & $0.749^{a}$ \\
$y1_{v}=y2_{v}$                    & $0.209^{a}$ \\
$L/(L_1+L_2)_V$                   & 0.6857 $\pm$0.0012 \\
r=R/a ($pole$)                   & 0.4547 $\pm$0.0005 &0.2767  $\pm$ 0.0014 \\
r=R/a ($side$)                   & 0.4895 $\pm$0.0007 &0.2892  $\pm$ 0.0017 \\
r=R/a ($back$)                   & 0.5187 $\pm$0.0008 &0.3267  $\pm$ 0.0032 \\
\hline
\end{tabular}
\end{center}
\end{table*}

\section{Conclusions}

Using the WDc and long term photometric observations of ROTSE-I J1643+25, we were able 
to determine its basic physical properties and establish that this system belongs to the group of W~UMa 
variables. With a mass ratio of $\sim$ 0.34, it presents an inclination of $\sim$ 84.6 degrees and a degree 
of over-contact of 11\%. The analysis of the data shows spots variations on the primary star, which imply 
strong activity on its surface.

\acknowledgments

\section*{Acknowledgments}
The authors are indebted to DGAPA (Universidad Nacional Aut\'onoma de M\'exico) support, PAPIIT projects IN111713. LFM, RM, TQ acknowledge the financial support from the UNAM under DGAPA grant PAPIIT IN 105115. The authors acknowledge the partial support from the DGAPA-UNAM  grant PAPIIT IN 106615.  This paper is based on observations acquired at the Observatorio Astron\'omico Nacional in the Sierra San Pedro M\'artir (OAN-SPM), Baja California, M\'exico. The authors would like to thank all the staff at the OAN-SPM for their invaluable help.

\end{document}